\documentclass[twocolumn,showpacs,preprintnumbers,amsmath,amssymb,times,pre]{revtex4}

\usepackage{graphicx}
\usepackage{dcolumn}
\usepackage{bm}
\usepackage{amsmath}
\usepackage{color}

\begin{document}

\title{Enhancement of cooperation through conformity-driven reproductive ability}

\author{Han-Xin Yang$^{1}$}\email{yanghanxin001@163.com}

\author{Lijun Tian$^{2}$}\email{buaatianlijun@163.com}

\affiliation{$^{1}$Department of Physics, Fuzhou University, Fuzhou
350116, China}

\affiliation{$^{2}$School of Economics and Management, Fuzhou
University, Fuzhou 350116, China}

\begin{abstract}
We propose a conformity-driven reproductive ability in which an
individual $i$ is more (less) likely to imitate a neighbor $j$'s
strategy if $j$'s strategy is majority (minority) in $i$'s
neighborhood. The results on the evolutionary spatial prisoner's
dilemma game show that, compared to homogeneous reproductive
ability, conformity-driven reproductive ability can greatly enhance
cooperation. This finding is robust with respect to different types
of network structures (including square lattice and scale-free
network) and to different ways of strategy updating (including
synchronous and asynchronous strategy updating).
\end{abstract}

\date{\today}

\pacs{02.50.Le, 87.23.Kg, 87.23.Ge}

\maketitle

\section{Introduction}

To understand the emergence of cooperative behavior among selfish
individuals, researchers have considered various mechanisms, such as
network reciprocity~\cite{1,2}, voluntary participation~\cite{3,4},
aspiration~\cite{5,6,7}, social diversity~\cite{9,10},
migration~\cite{11,12,13}, chaotic payoff variations~\cite{14},
extortion~\cite{15,16,17}, punishment~\cite{the1,the2,the3,the4},
and so on.

The reproductive ability, also is known as the teaching ability or
the learning ability, has been extensive studied in the evolutionary
games~\cite{a1,a2,a3,a4,a5,a6}.  Szolnoki $et$ $al$ proposed an
inhomogeneous teaching ability in which the probability that the
individual $i$ adopts a randomly chosen neighbor $j$'s strategy
depends on the payoff difference and a two-value pre-factor $\omega$
characterizes the teaching ability of neighbor $j$~\cite{a1}. They
found that inhomogeneous teaching ability can promote cooperation
for the prisoner's dilemma games in lattices~\cite{a1} and complex
networks~\cite{a2}. Guan $et$ $al$ found that the introduction of
the inhomogeneous activity of teaching of individuals can remarkably
promote cooperation in spatial public goods games~\cite{a3}.
Szolnoki and Perc defined the teaching ability of a node $i$ as its
collective influence which is the product of its reduced degree and
the total reduced degree of all $j$ nodes at a hierarchial depth
$\ell$ from node $i$~\cite{a4}. It was found that there exists an
optimal hierarchical depth for the determination of collective
influence that favors cooperation. Chen $et$ $al$ proposed an
inhomogeneous learning ability in which the two-value pre-factor
$\omega$ characterizes the strength of the individual $i$'s own
learning activity~\cite{a5}. They found that appropriate
intermediate levels of learning activity can promote or sustain
cooperation for the prisoner's dilemma games in small-world networks
and scale-free networks. Wu $et$ $al$ discovered that cooperation on
square lattices is promoted (inhibited) in the case of synchronous
(asynchronous) strategy updating, if heterogeneous learning ability
is considered~\cite{a6}.

In many real-life situations, an individual tends to follow the
majority in behavior or opinion within the interaction range.
Recently, the consideration of conformity has attracted much
attention in the study of evolutionary games. Szolnoki and Perc
designated a fraction of population as being driven by conformity
rather than payoff maximization~\cite{inter1,inter2}. These
conformists simply adopt whichever strategy is most common within
their interaction range at any given time, regardless of the
expected payoff. They showed that an appropriate fraction of
conformists within the population introduces an effective surface
tension around cooperative clusters and ensures smooth interfaces
between different strategy domains.

Motivated by the work of Szolnoki and Perc, we propose a
conformity-driven reproductive ability in which the probability that
the individual $i$ adopts a randomly chosen neighbor $j$'s strategy
depends on the payoff difference and a pre-factor $\omega_{ij}$
characterizing the popularity of $j$'s strategy among $i$'s
neighbors. The value of $\omega_{ij}$ is above (below) 0.5 if $j$'s
strategy is majority (minority) in $i$'s neighborhood. Different
from previous works of the teaching ability, the pre-factor in our
model is determined not only by $j$ but also by $i$'s other
neighbors.

\section{Model}\label{sec:model}

Our model is described as follows.

Player $x$ can take one of two strategies: cooperation or defection,
which are described by
\begin{equation}\label{1}
s_{x} =\left(
     \begin{array}{c}
       1 \\
       0 \\
     \end{array}
   \right)\mathrm{or}\left(
     \begin{array}{c}
       0 \\
       1 \\
     \end{array}
   \right),
\end{equation}
respectively. At each time step, each individual plays the
prisoner's dilemma game with its nearest neighbors. An individual
will punish the neighbors that hold different strategies. The
accumulated payoff of player $x$ can thus be expressed as
\begin{equation} \label{2}
P_{x}=\sum_{y\in \Omega_{x}}s_{x}^{T}Ms_{y},
\end{equation}
where the sum runs over the nearest neighbor set $\Omega_{x}$ of
player $x$ and $M$ is the rescaled payoff matrix given by
\begin{equation}\label{3}
M=\left(
    \begin{array}{cc}
      1 & 0 \\
      b & 0 \\
    \end{array}
  \right).
\end{equation}
Here the parameter $b (>1)$ denotes the temptation to defect.

Initially, cooperators and defectors are randomly distributed with
the equal probability 0.5. Players asynchronously update their
strategies in a random sequential
order~\cite{random1,random2,random3}. Firstly, an individual $i$ is
randomly selected who obtains the payoff $P_{i}$ according to the
above equations. Next, individual $i$ chooses one of its nearest
neighbors at random, and the chosen neighbor $j$ also acquires its
payoff $P_{j}$. Finally, individual $i$ adopts the neighbor $j$'s
strategy with the probability~\cite{a1}:
\begin{equation}\label{4}
W(s_{i}\leftarrow s_{j})=\omega_{ij}\frac{1}{1+\exp[(P_i-P_j)/K]},
\end{equation}
where $K$ characterizes the noise introduced to permit irrational
choices and $\omega_{ij}$ characterizes the ability that $j$
transfers its strategy to $i$.

We define the reproductive ability $\omega_{ij}$ as:
\begin{equation}\label{5}
\omega_{ij}=\frac{1}{1+\exp[(k_i/2-N_{s_j})/H]},
\end{equation}
where $N_{s_j}$ is the number of players adopting strategy $s_{j}$
within the interaction range of player $i$ (including $j$ itself),
$k_i$ is the degree of player $i$, and $H (>0)$ represents the
steepness of the function. The more popular is $j$'s strategy in
$i$'s neighborhood, the higher is the reproductive ability of $j$.
For $H=\infty$, the reproductive ability is a constant equaling 0.5.
In this situation, our model is restored to the original homogeneous
ability. Conversely, for $H=0$ the reproductive ability becomes
steplike so that $i$ refuses to adopt strategy $s_{j}$ if this
strategy is minority in $i$'s neighborhood.

\section{Results}\label{sec: results}

\begin{figure}
\begin{center}
\scalebox{0.4}[0.4]{\includegraphics{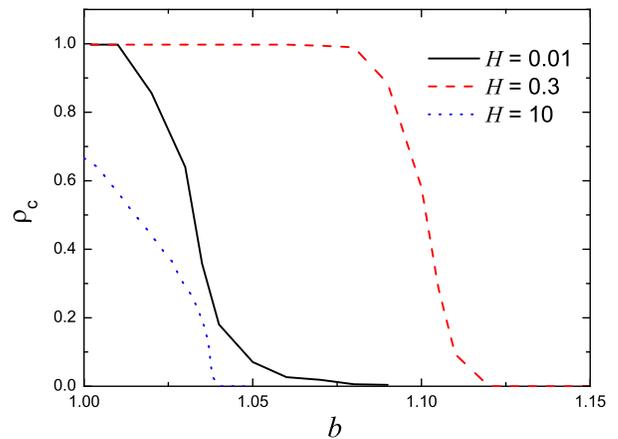}} \caption{(Color
online) The fraction of cooperators $\rho_{c}$ as a function of the
temptation to defect $b$ for different values of the steepness
parameter $H$. For each value of $H$, $\rho_{c}$ decreases to 0 as
$b$ increases. For small values of $H$ (e.g., $H=0.01$ or $H=0.3$),
cooperators can occupy the whole system when $b$ is small. However,
for large values of $\alpha$ (e.g., $\alpha=10$), full cooperation
cannot reach even $b=1$.} \label{fig1}
\end{center}
\end{figure}

\begin{figure}
\begin{center}
\scalebox{0.4}[0.4]{\includegraphics{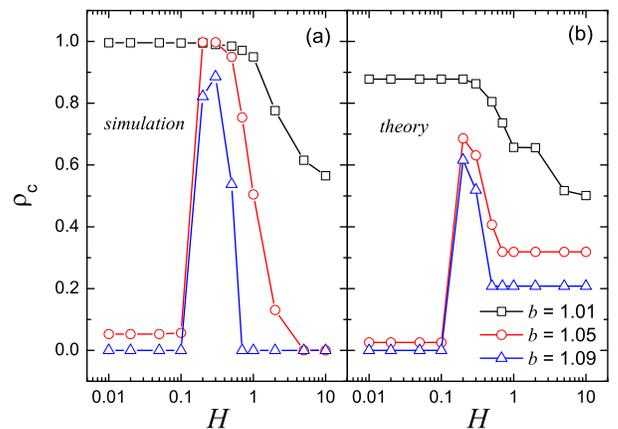}} \caption{(Color
online) The fraction of cooperators $\rho_{c}$ as a function of the
steepness parameter $H$ for different values of the temptation to
defect $b$. The results from (a) simulation and (b) theoretical
analysis, respectively.} \label{fig2}
\end{center}
\end{figure}

Following the previous studies~\cite{noise1,noise2}, we set the
noise level to be $K = 0.1$. The key quantity for characterizing the
cooperative behavior of the system is the fraction of cooperators
$\rho_{c}$ in the steady state.  In the following simulation,
$\rho_{c}$ is obtained by averaging over the last $10^{3}$ Monte
Carlo steps (MCS) of the entire $10^{5}$ MCS. Each MCS consists of
on average one strategy-updating event for all individuals. Each
data is obtained by averaging over 100 different realizations.
Unless otherwise specified, all our simulations are performed in a
$100 \times 100$ square lattice with the periodic boundary
condition.

\begin{figure*}
\begin{center}
\scalebox{0.78}[0.78]{\includegraphics{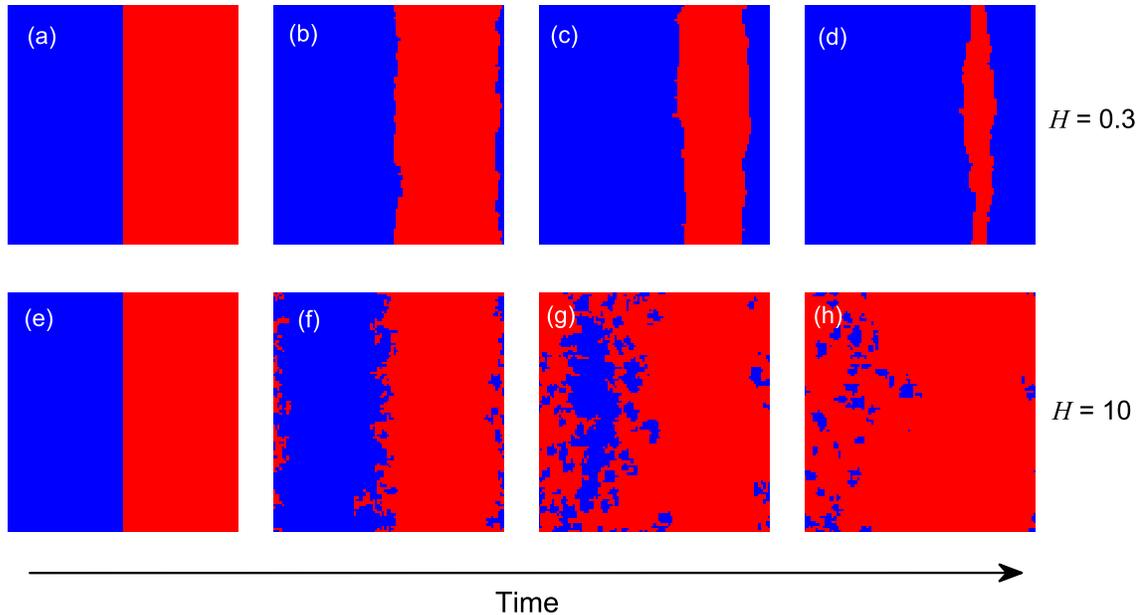}} \caption{(Color
online) Snapshots of typical distributions of cooperators (blue) and
defectors (red) at different time steps. Initially, we set
cooperators (defectors) in the left (right) half of square lattices.
The temptation to defect $b=1.05$. The steepness parameter $H$ is
$H=0.3$ for (a)-(d) and $H=10$ for (e)-(h). For $H=0.3$, the
cooperator cluster continually expands and the boundary between the
two competing clusters keeps smooth during the whole evolution. For
$H=10$, the defector cluster continually expands and the boundary
becomes littery as time evolves.} \label{fig3}
\end{center}
\end{figure*}

\begin{figure}[htbp]
\begin{center}
\scalebox{0.37}[0.37]{\includegraphics{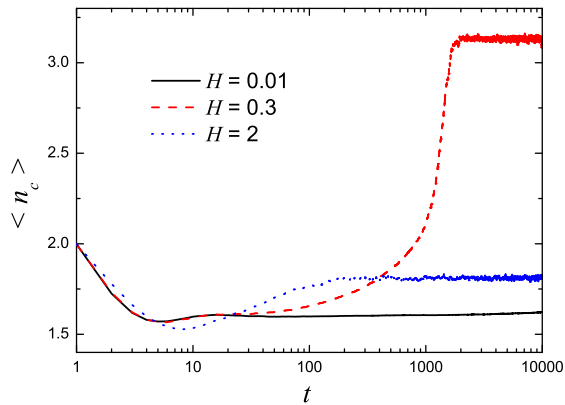}} \caption{(Color
online) The time evolution of the average number of cooperative
neighbors $\langle n_{c} \rangle$ for players along the interfaces
separating domains of cooperators and defectors. We define a player
along the the interface as the one who has at least one neighbor
with the opposite strategy. The temptation to defect $b=1.05$. For
each value of the steepness parameter $H$, $\langle n_{c} \rangle$
firstly decreases and then increases as time evolves. } \label{fig4}
\end{center}
\end{figure}

Figure~\ref{fig1} shows the fraction of cooperators $\rho_{c}$ as a
function of the temptation to defect $b$ for different values of the
steepness parameter $H$. From Fig.~\ref{fig1}, one can see that, for
each value of $H$, $\rho_{c}$ decreases to 0 as $H$ increases. For
small values of $H$ (e.g., $H=0.01$ or $H=0.3$), full cooperation
can reach when $b$ is below a threshold value. However, for large
values of $H$ (e.g., $H=10$), cooperators cannot take over the whole
system even $b=1$.

Figure~\ref{fig2} shows the fraction of cooperators $\rho_{c}$ as a
function of the steepness parameter $H$ for different values of the
temptation to defect $b$. We see that, for relatively small values
of $b$ (e.g., $b = 1.01$),  $\rho_{c}$ decreases as $H$ increases.
However, for larger values of $b$ (e.g., $b = 1.05$ or 1.09), there
exists an optimal value of $H$ (about 0.2) leading to the highest
cooperation level. The dependence of $\rho_{c}$ on $H$ can be
qualitatively predicted analytically through a pair-approximation
analysis~\cite{pair1,pair2}, the results of which are shown in Fig.
~\ref{fig2}(b).

To intuitively understand why the moderate value of $H$ that can
best enhance cooperation, we plot spatial strategy distributions as
time evolves for different values of $H$ when the temptation to
defect $b=1.05$. Initially we set a giant cooperator (defector)
cluster in the left (right) half of square lattices. From
Figs.~\ref{fig2}(a)-(d), one can see that for the moderate value of
$H$ (e.g., $H=0.3$), the cooperator cluster continually expands
while the defector cluster gradually shrinks. Note that for $H=0.3$,
the boundary between the two competing clusters remains smooth
during the whole evolution. However, for the large value of $H$
(e.g., $H=10$), the defector cluster gradually invade the cooperator
cluster and the original one big cooperator cluster is divided into
some small clusters [see Figs.~\ref{fig2}(e)-(h)]. For $H=10$, the
interfaces separating domains of cooperators and defectors become
littery. As pointed out in Ref.~\cite{border1,border2}, noisy
borders are beneficial for defectors, while straight domain walls
help cooperators to spread. For the very small value of $H$ (e.g.,
$H=0.01$), the cooperator and defector clusters keep almost
unchanged (the results are not shown here).

\begin{figure*}
\begin{center}
\scalebox{0.8}[0.8]{\includegraphics{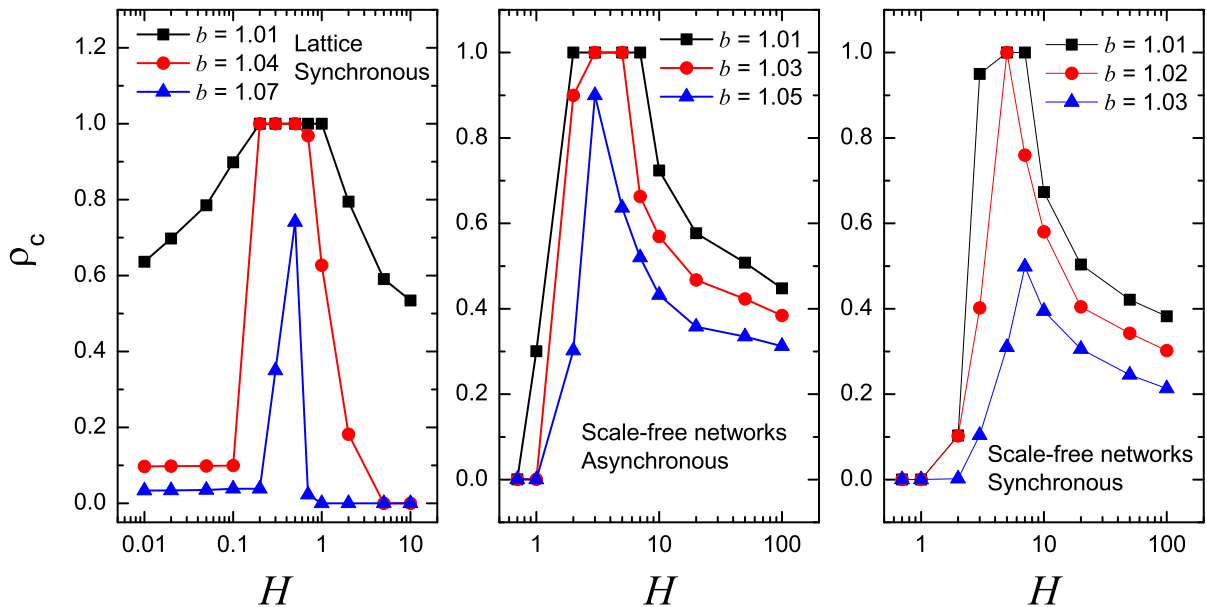}} \caption{(Color
online) The fraction of cooperators $\rho_{c}$ as a function of the
steepness parameter $H$ for different values of the temptation to
defect $b$ under different types of networks and different kinds of
updating rules. Left panel: for square lattices and synchronous
updating rule. Middle panel: for scale-free networks and
asynchronous updating rule. Right panel: for scale-free networks and
synchronous updating rule. The network size is set to be 10000 and
the average degree of the network is 4. Note that for scale-free
networks, we use degree-normalized payoffs. } \label{fig5}
\end{center}
\end{figure*}

Next, we study the average number of cooperative neighbors $\langle
n_{c} \rangle$ for players along the interfaces separating domains
of cooperators and defectors. A player is along the interface if it
has at least one neighbor with the opposite strategy.
Figure~\ref{fig4} shows the time evolution of $\langle n_{c}
\rangle$ for different values of the steepness parameter $H$ when
the temptation to defect $b=1.05$. One can see that initially
$\langle n_{c} \rangle$ decreases from 2 to about 1.6, and then
increases to a stable value. For the small value of $H$ (e.g.,
$H=0.01$) and the large value of $H$ (e.g., $H=2$), the final value
of $\langle n_{c} \rangle$ is below 2. However, for the moderate
value of $H$ (e.g., $H=0.3$), $\langle n_{c} \rangle$ finally
reaches 3.1. Once the value of $\langle n_{c} \rangle$ exceeds 2,
cooperation becomes the majority strategy in a player's
neighborhood. In this case, the conformity-driven reproductive
ability is beneficial for the expansion of cooperator clusters.

In all the above studies, we use square lattices and asynchronous
strategy updating. In fact, our finding that the moderate value of
the steepness parameter $H$ can best promote cooperation is robust
with respect to different kinds of network structures and different
ways of strategy updating. Since square lattice is homogeneous
interaction networks, it is interesting for us to consider
heterogeneous interaction networks. We use the famous
Barab\'{a}si-Albert scale-free networks to construct heterogeneous
interaction~\cite{BA}. In the asynchronous updating rule, at each
time only a randomly selected player is allowed update its strategy.
While in synchronous updating rule, at each time all players update
their strategies. We consider three cases: square lattices with
synchronous strategy updating, scale-free networks with asynchronous
strategy updating and scale-free networks with synchronous strategy
updating. From Fig.~\ref{fig5}, one can see that for all these
cases, the cooperation level reaches the highest at the moderate
value of $H$ when the temptation to defect $b$ is fixed.

\section{Conclusions and discussions}\label{sec: conclusion}

To summarize, we have proposed a conformity-driven reproductive
ability in which the probability that a player $i$ adopts a neighbor
$j$'s strategy depends on their payoff difference and a pre-factor
$\omega_{ij}$ characterizing the popularity of $j$'s strategy among
$i$'s neighbors. The value of $\omega_{ij}$ increases with the
number of $i$'s neighbors holding the same strategy with $j$. Both
numerical and theoretical results show that, the cooperation level
of the spatial prisoner's dilemma game can be greatly enhanced by
moderately increasing the teaching ability of the neighbor with the
majority strategy in the local community. In the case of the
conformity-driven reproductive ability, the borders of cooperator
clusters become smooth, thus cooperators along the borders can get
more help to resist the invasion of defectors. Note that the concept
of conformity is widely existent in opinion
dynamics~\cite{majority1,majority2}. We hope our work can attract
more interest in the study of the heterogeneous reproductive ability
based on the opinion dynamics.

\begin{acknowledgments}
This work was supported by the National Natural Science Foundation
of China (Grants Nos. 61403083, 71301028 and 71671044), and
Excellent Youth Science Foundation of Fujian Province (Grant No.
2016J06017).
\end{acknowledgments}

\end{document}